\documentclass{aa}
\usepackage{txfonts}
\usepackage{times,psfig}

\def\ltsima{$\; \buildrel < \over \sim \;$}
\def\lsim{\lower.5ex\hbox{\ltsima}}
\def\gtsima{$\; \buildrel > \over \sim \;$}
\def\gsim{\lower.5ex\hbox{\gtsima}}

\newcommand{\be}{\begin{equation}}
\newcommand{\en}{\end{equation}}
\newcommand{\ergs}{\rm \ erg \; s^{-1}}

\def\cmdue {\rm \ cm^{-2}}

\def\msole {~M_{\odot}}

\begin{document}

\title{XMM-Newton observations of two transient millisecond X--ray pulsars in
quiescence}  

\author{S.~Campana\inst{1}, N. Ferrari\inst{1,2},
L. Stella\inst{3}, G.L. Israel\inst{3}}

\authorrunning{S. Campana}

\titlerunning{XMM-Newton observations of two transient millisecond pulsars}

\offprints{Sergio Campana, campana@merate.mi.astro.it}

\institute{INAF-Osservatorio Astronomico di Brera, Via Bianchi 46, I--23807
Merate (Lc), Italy
\and
Universit\`a degli Studi di Milano, Via Celoria 16, I--20133
Milano, Italy
\and
INAF-Osservatorio Astronomico di Roma,
Via Frascati 33, I--00040 Monteporzio Catone (Roma), Italy}

\date{received, accepted}

\abstract{
We report on XMM-Newton observations of two X--ray transient millisecond
pulsars (XRTMSPs). We detected XTE J0929--314 with an unabsorbed luminosity of
$\sim 7\times 10^{31}\ergs$ (0.5--10 keV) at a fiducial distance of 10 kpc. The
quiescent spectrum is consistent with a simple power law spectrum. The upper
limit on the flux from a cooling neutron star atmosphere is about $20\%$ of
the total flux. XTE J1807--294 instead was not detected. We can put 
an upper limit on the source quiescent 0.5--10 keV unabsorbed luminosity
$\lsim 4\times10^{31}\ergs$ at 8 kpc. These observations strenghten the idea
that XRTMSPs have quiescent luminosities significantly lower than classical
neutron star transients.

}

\maketitle

\section{Introduction}

X--ray transient millisecond pulsars (XRTMSPs) provided the (long-sought) direct
evidence that the neutron stars in Low Mass X--ray Binaries are spinning fast
and possess a sizable magnetosphere. Up to now six sources of this type have been
discovered thanks to the Rossi X--ray Timing Explorer (RXTE): SAX J1808--3654
(Wijnands \& van der Klis 1998), XTE J1751--305 (Markwardt et al. 2002), XTE
J0929--314 (Remillard, Swank \& Strohmayer 2002), XTE J1807--294 (Markwardt,
Smith \& Swank 2003; Campana et al. 2003), XTE J1814--338 (Markwardt, Juda \& 
Swank 2003) and, very recently, IGR J00291+5934 (Markwardt, Swank \&
Strohmayer 2004).

XRTMSPs belong to the so called neutron star Soft X--ray transient (SXRT)
class and form a distinct subgroup. SXRTs show bright outbursts with peak
X--ray luminosities in the $10^{36}-10^{38}\ergs$ range. SXRTs when in quiescence
usually have a luminosity in the $10^{32}-10^{33}\ergs$ range (for a review
see Campana et al. 1998a). Their quiescent spectra can be described by a soft
component (modelled as a neutron star atmosphere spectrum from the entire
surface) plus a power law tail which is present in a good number of 
sources\footnote{This power law tail is not present at all (even with very
tight limits) in quiescent Low Mass X--ray Binaries found in increasing
numbers in globular clusters (Heincke et al. 2003).} 
and can account for up to $\sim 70\%$ of the 0.5--10 keV flux. One exception
is represented by the transient EXO 1745--248 in the globular cluster
Terzan 5. The quiescent state of this source is bright (a few $10^{33}\ergs$)
and its spectrum can be modelled by a power law tail only (Wijnands et 
al. 2005a). This is just one source in about a dozen of SXRTs.  

\begin{figure*}[htbp]
\begin{center}
\psfig{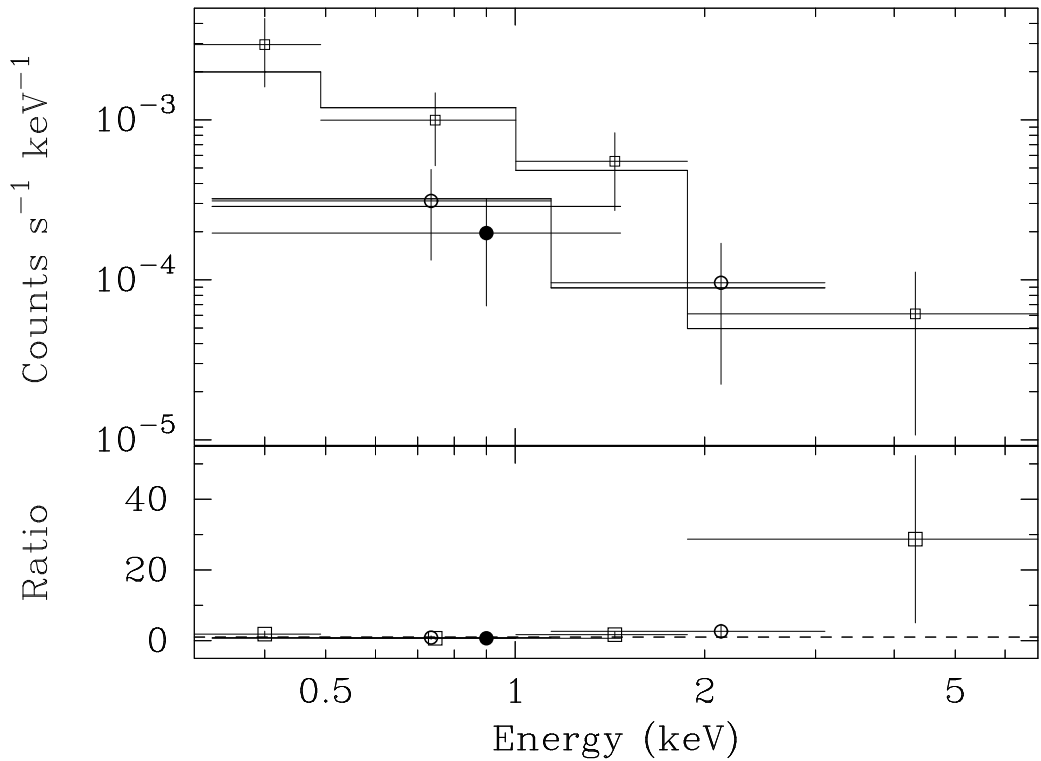}
\end{center}
\caption{XMM-Newton spectrum of XTE J0929--314. In the upper panel data
along with the best fit power law spectrum are shown: pn data are marked with a
open squares, MOS1 data with open circles and MOS2 data with filled
circles. In the bottom panel the ratio between the best fit NSA model and the
data is plotted. It is apparent that the model is underestimating the
data at energies larger than $\sim 2$ keV.}
\label{curve}
\end{figure*}

XRTMSPs display significant differences. Their outbursts are a factor of
10--100 fainter than those of most SXRTs. When in quiescence, XRTMSPs are also
fainter. SAX J1808.4--3658 was the first XRTMSP observed in
quiescence. XMM-Newton observations allowed us to settle its quiescent
spectrum on a firm basis. The spectrum could be modelled with a single power
law ($\Gamma=1.5^{+0.2}_{-0.3}$, $90\%$ confidence level, Campana et
al. 2002). For a distance of 2.5 kpc (in't Zand et al. 2001) SAX J1808.4--3658
is underluminous by a factor of two with respect to SXRTs in quiescence
($L\sim 5\times10^{31}\ergs$).
Recently, Chandra observed two other millisecond X--ray pulsars, XTE J0929--314
and XTE J1751--305 (Wijnands et al. 2005b). XTE J0929--314 was detected with
22 photons only, whereas XTE J1751--305 was not detected (upper limit of 5
photons).
A spectral analysis of the XTE J0929--314 data, despite poor statistics,
suggested a simple power law spectrum ($\Gamma=1.8^{+0.6}_{-0.5}$, $90\%$ confidence
limit), giving a low 0.5--10 keV luminosity of $7^{+5}_{-2}\times
10^{31}\,(d/10 {\rm \,kpc})^2\ergs$ (note, however, that the source distance
is not yet known). An upper limit of $\sim 30\%$ on the flux of a soft
component could be set. For XTE J1751--305 an upper limit of $(0.2-2)\times
10^{32}\,(d/8 {\rm \,kpc})^2\ergs$ on the 0.5--10 keV luminosity was determined,
depending on the assumed spectral parameters (Wijnands et al. 2005b). {Taken
at face value,} these results confirmed that the quiescent counterparts of
XRTMSPs are dimmer than those of standard SXRT sources.  

Here we report on XMM-Newton observations of XTE J0929--314 (section 2) and XTE
J1814--338 (section 3). Our conclusions are described in section 4.

\section{XMM-Newton observation of XTE J0929--314}

XMM-Newton observed XTE J0929--314 on May 4 2004 for 30 ks. The entire
observation was plagued by soft proton flares and no usable data were
obtained. A second observation took place on June 12 2004 for 15.9 ks. The
thin filter was used for all three EPIC cameras. We filtered out light
background flares for total rates (0.2--15 keV) less than $3.5$ [8] c s$^{-1}$
for the MOS [pn] cameras, resulting in 15 [11] ks of good time intervals. The
source was barely visible in the pn image. We extracted the pn source spectrum
(circular region with $r=13''$) and the background spectrum (three circular
regions around the source on the same CCD with $r=30''$). For the MOS cameras
we did the same with a source extraction region of $r=13''$ and four
background regions of $90''$ around the source. We obtained about 10 (35) counts
for MOS (pn) camera in the 0.3--10 keV range, of which about $40\%$ are 
background events. A search at the known transient position provided a detection
with a signal to noise ratio of $\sim 4$.

We then fit together the three spectra. Given the low number of counts, we
binned the MOS data to 5 counts per energy channel and the pn data to 8 counts
per bin. Given the small number of counts we adopted the Cash statistics (Cash
1979). A soft component model (neutron star atmosphere or black body) did not
provide a satisfactory description of the data (see Table 1). We simulated
10,000 spectra (with parameter values drawn from a Gaussian distribution
centered on the best fit with sigma from the covariance matrix): about
$25-40\%$ of them had a better C-statistics than the actual data. We thus fit 
the data with a single power law obtaining better results. In this case only
$\sim 1\%$ of the simulated data had a better C-statistics than the real
data. The power law photon index was constrained to be
$\Gamma=2.0^{+1.9}_{-0.7}$. A similarly good fit (goodness of $\sim 1\%$) could
be obtained with the canonical model for neutron star transients in
quiescence, i.e. a neutron star atmosphere model plus a power law tail. To
derive an upper limit we freeze the best fit power law model and choose the
maximal soft component that can be accomodated within the power law fit,
degreding the spectral fit to a $10\%$ acceptance with the Cash statistics.
The upper limit to the contribution of the soft component in the 0.3--10 keV
energy band is $\sim 20\%$. The neutron star atmosphere temperature is very
soft  with $T=35-55$ eV (depending on the selected normalization) and its
bolometric flux amounts to $<2\times 10^{-15}\ergs\cmdue$, corresponding to
luminosity $<3\times 10^{31}\ergs$ at 10 kpc.  

Finally, we searched for flux and spectral variations across the observation as
tentatively suggested by Wijnands et al. (2005b) but we did not find any significant
variation. 

\begin{table*}
\caption{XTE J0929--314 spectral fits.}
\begin{center}
\begin{tabular}{cccccc}
\hline
Model    & $N_H$        & Temp./Power law& Flux$^+$          & C-stat.$^*$ & Goodness$^\dag$\\
         & $10^{21}\cmdue$&  (eV)          & ($10^{-15}$ cgs)&($chi^2_{\rm red}$)&  \\
\hline
NSA      & $<3.3$       &$T=76^{+16}_{-31}$  & 3.8 (6.1)    & 6.2 (1.3)  & $41.6\%$ \\
Blackbody& $<2.1$       &$T=267^{+102}_{-125}$& 2.9 (3.3)    & 5.1 (1.0)  & $25.1\%$ \\
Power law& $<2.6$       &$\Gamma=2.0^{+1.9}_{-0.7}$&5.4 (6.3)& 1.3 (0.3)  &\ $0.5\%$ \\
Pow+NSA  & $<2.1$       &$T=33^{+30}_{-25}$   & 6.6 (7.8)    & 0.8 (0.4)  &\ $0.3\%$ \\
         &             &$\Gamma=1.6^{+1.7}_{-2.0}$& $>95\%$ &            & \\
\hline
\end{tabular}
\end{center}

All errors are $90\%$ confidence level, obtained with $\Delta \chi^2=2.71$.\\
Soft component models have a normalization free to vary within the distance
range 5--15 kpc. 

$^+$ Fluxes are unabsorbed. The first value refers to the 0.5--10 keV energy
range, the value in parenthesis to the 0.3--10 keV range. In the last raw the
lower limit refers to the contribution of the power law to the total flux.

$^*$ C-statistics value for 7 energy bins (in parenthesis is reported the
reduced $chi^2$ for the same fit).

$^\dag$ Percentage of simulated data (Gaussian distribution) which gives a better
C-statistics than real data.

\label{template}
\end{table*}

\section{XMM-Newton observation of XTE J1807--294}

XMM-Newton observed XTE J1807--294 on September 8 2004 for 39 ks. The thin
filter was used for all the three EPIC cameras. During the observation,
several soft proton flares occurred. We filtered MOS [pn] total rate
(0.2--15 keV) with at a limiting rate of $5$ [10] c s$^{-1}$ for the MOS [pn]
cameras, resulting in 32 [5] ks of good time intervals.
XTE J1807--294 was not detected in the EPIC images. A $3\,\sigma$ upper limit
of $8\times 10^{-4}$ c s$^{-1}$ is derived from the 0.5--10 keV pn image. This
rate can be translated into a flux assuming a spectral model. We consider the
column density observed in outburst $N_H=5\times 10^{21}\cmdue$ (Campana et
al. 2003). For a power law with photon index in the $1.5-2$ range 
a limit on the (unabsorbed) flux of $\sim 5\times10^{-15}\ergs\cmdue$ can
be derived; for a black body model with temperatures in the 0.1--0.3 keV
range a limit of $\sim 3\times10^{-15}\ergs\cmdue$ can be derived. The
distance to XTE J1807--294 is presently unknown. A scale distance of 8 kpc is
assumed. At this distance the flux limit implies an upper limit on the 0.5--10
keV luminosity of $\sim 4\times 10^{31}\ergs$. 

\begin{table}
\caption{Quiescent 0.5--10 keV luminosities of neutron star transients.}
\begin{center}
\begin{tabular}{cccc}
Name             & Distance (kpc) & $\log L$ & Ref.\\
\hline
SAX J1808.4--3658& $3.15\pm0.45$     & 31.9 &1,2\\
XTE J0929--314   & $10\pm5     $     & 31.8 &3,4\\
IGR J00291+5934  & $3\pm3      $     & 31.7 &5\\
XTE J1807--294   & $8\pm5      $     & $<31.6$ &4\\
XTE J1751--305   & $8\pm5      $     & $<31.5$ &3\\
\hline
Cen X-4          & $1.4\pm0.3  $     & 32.7 &6,7\\
4U 1608--52      & $3.3\pm0.5  $     & 33.2 &8,2\\
MXB 1659--29     & $9.85\pm1.45$     & 32.5 &9,2\\
XTE J1709--267   & $10\pm2     $     & 33.3 &10\\
KS 1731--260     & $6.2\pm0.9  $     & 32.4 &11,2\\
SAX J1810.8--2609& $5.95\pm0.85$     & 32.2 &12,2\\
Aql X-1          & $5.15\pm0.75$     & 33.4 &13,2\\
XTE J2123-058    & $18.35\pm2.65$    & 32.7 &14,2\\
Terzan 5         & $8.7\pm3    $     & 33.3 &15,2\\
\hline
\end{tabular}
\end{center}
\label{distance}
References -- 1: Campana et al. 2002; 2: Jonker \& Nelemans 2004; 3: Wijnands et
al. 2005b; 4: this work; 5: Jonker et al. 2005; 6: Campana et al. 2004; 7:
Gonz\'alez Hernandez et al. 2005; 8: Asai et al. 1996; 9: Wijnands et
al. 2004; 10: Jonker et al. 2004a; 11: Wijnands et al. 2001; 12: Jonker et
al. 2004b; 13: Campana et al. 1998b; 14: Tomsick et al. 2004; 15: Wijnands et
al. 2005a. 
\end{table}

\section{Conclusions}

It is becoming apparent that the five XRTMSPs are different from the great
majority of SXRTs not only for their outburst properties (periodicities, faint
outbursts) but also for their properties in quiescence. 

The first object of this class, SAX J1808.4--3658, has been rapidly recognized
to have peculiar properties in quiescence, namely the low X--ray luminosity
and a simple power law spectrum (Campana et al. 2002).
Confirmation that XRTMSP are peculiar came from Chandra observations of XTE
J0929--314 and XTE J1751--305 (Wijnands et al. 2005b). XTE J0929--314 was
detected at a very low level of $\sim 7\times 10^{31}\ergs$ (0.5--10 keV) for
a fiducial distance of 10 kpc. The quiescent spectrum is consistent with a
simple power law spectrum. The upper limit on the flux from a cooling neutron
star atmosphere is about $30\%$ of the total flux. XTE 1751--305 was not
detected with an upper limit of $\sim 2\times 10^{32}\ergs$ (at 8 kpc, Wijnands et
al. 2005b). 

We observed XTE J0929--314 with XMM-Newton. Our spectrum is consistent with
the Chandra one. Given the better response at high energies we can put a
stronger limit of $\lsim 20\%$ on the presence of a neutron star atmosphere
component (consistent with the emission from the entire surface in the 5--15
kpc distance range). The temperature of the putative soft component is
extremely soft ($\sim 35-55$ eV) so that the bolometric luminosity would be $\sim
3\times 10^{31}\ergs$, at the most. This limit is tighter than the one derived
by Wijnands et al. (2005b). Actually, under the hypothesis that deep crustal
heating (DCH) is at work (and there are no reasons to beleive it should
not), from the observed quiescent luminosity (or the limit on it) one can gain
insight on the accretion history of the source (Brown et al. 1998). Wijnands
et al. (2005b) estimated a mean accretion 
powered flux of $F_{\rm acc}=3.5 \times 10^{-11}\ergs\cmdue$ (for the $\sim 9$
years of RXTE monitoring). Based on DCH theory, the estimate of the mean
accretion-powered flux provides a prediction for the quiescent flux $F_{\rm
q}=F_{\rm acc}/135=2.6\times 10^{-13}\ergs\cmdue$ (Brown et al. 1998). This is
much higher than observed. In order to match our observed upper limit we have
to require a much longer quiescent period than the 9 yr of RXTE
observations. Using an outburst duration of $t_{\rm o}\sim 73$ d (Galloway et
al. 2002) and average flux during outburst $F_{\rm o}\sim 1.6\times 
10^{-9}\ergs\cmdue$, we can estimate the recurrence time  $t_{\rm q}$ from
$F{\rm q}\sim {{t_{\rm o}}\over{t_{\rm o}+t_{\rm q}}}\times {{F_{\rm
o}}\over{135}}$. We then derive $t_{\rm q} \sim 780$ yr. This recurrence time
is long, even for the most extreme version of disk instability
models (Lasota 2001). 

Following Wijnands et al. (2005b) we can also estimate the expected quiescent
luminosity due to DCH based on a mass transfer rate driven by
gravitational radiation. We estimated a luminosity of $L_{\rm DCH-GW}\sim
3\times10^{32}\,M_{NS}^{2/3}\,M_{c}^2\,Q\ergs$ (assuming a neutron star and a
companion mass $M_{NS}=1.4\msole$ and $M_{c}=0.008\msole$, respectively, and the
amount of heat deposited in the crust per accreted nucleon $Q=1.45$ MeV). 
The source distance must be $\sim 30$ kpc to be consistent with this limit.

As already suggested for SAX J1808.4--3658 (Campana et al. 2002) there is an
easy explanation for the absence of a soft component in the quiescent spectrum
of XTE J0929--314 as well. The theory of deep crustal heating
consider only a standard cooling scenario, additional cooling results
in lower luminosities. A simple and well-known solution is when the direct
Urca process is at work in the neutron star core and neutrino 
cooling affects the neutron star thermal evolution (Colpi et
al. 2001, Yakovlev \& Pethick 2004). This can occur only for massive neutron
stars with masses higher than $\sim 1.7-1.8\msole$. 
On the other side the power law tail can result either from low level
accretion or from the interaction between the relativistic pulsar wind of a
turned-on radio pulsar and matter outflowing from the companion (for a
detailed discussion see e.g. Campana \& Stella 2000).

In the case of XTE J1807--294 our XMM-Newton observation allowed us only to put
an upper limit on the source quiescent 0.5--10 keV unabsorbed flux of $\lsim 
5\times10^{-15}\ergs\cmdue$ or luminosity $\lsim 4\times10^{31}\ergs$ at 8
kpc. This limits testifies once more that the quiescent emission of
XRTMSP is fainter than classical SXRTs.
We can also test the DCH predictions on XTE J1807--294 given its flux history
(Markwardt, Smith \& Swank 2003) in the last few years. Integrating the flux history
provided by RXTE pointed data we obtain a 2--10 keV fluence of $\sim 4\times
10^{-3}$ erg cm$^{-2}$ during the only outburst detected. A bolometric
correction factor for the extrapolation of the spectrum to 
the 0.1--60 keV energy range amounts to $\sim 2.3$, taking the XMM-Newton
spectrum in outburst (Campana et al. 2003). If no other outbursts occurred
during the RXTE lifetime, we can derive a mean accretion flux of $\lsim
3\times 10^{-11}\ergs\cmdue$. This turns into a predicted quiescent flux of 
$F_{\rm q}\lsim 2\times 10^{-13}\ergs\cmdue$, much higher than that observed. 
Taking the upper limit for black body emission derived above, this implies
having a recurrence time of $\gsim 7$ yr, taking a short outburst duration of
$\sim 40$ d (corresponding to the time interval during which we have published
data).

The quiescent luminosities of XRTMSPs we obtained confirmed once more that XTE
J0929--314 and XTE J1807--294 are fainter than classical SXRTs. We
considered the 9 SXRT with known quiescent X--ray luminosities and good
distance estimates (Jonker \& Nelemans 2005, see Table 2) and compared them
with the 5 XRTMSPs (inluding the recently discovered IGR J00291+5934,
Jonker et al. 2005). In order to account for the uncertainties in the
source distances, we run a MonteCarlo simulation with a random distance within
the quoted range in Table 1, and compare the luminosities of the two samples
with a Kolmogorov-Smirov test. The (logaritmic) mean chance probability that 
the two distributions are drawn from the same parent population is $\sim
1.4\%$. Clearly this test cannot be regarded as conclusive given, e.g., 
the presence of upper limits. Despite this, it provides a clear indication of
the difference between the two populations at $\sim 2.5\sigma$ confidence level.

The XRTMSP population in quiescence represent a new challenge for the 
physics of neutron stars. These sources are extremely faint in quiescence, they can be
easily detected by Chandra but only poor spectroscopic studies can be carried
out by exploiting the XMM-Newton higher throughput, calling for future
observations with satellites such as Constellation-X and XEUS.


\begin{thebibliography}{}

\bibitem []  {}
Asai K., Dotani T., Mitsuda K., et al.  1996, PASJ 48 257 

\bibitem []  {}
Brown E.F., Bildstein L., Rutledge R.E. 1998, ApJ 504 L95

\bibitem []  {}
Campana S., Colpi M., Mereghetti S., Stella L., Tavani M. 1998a, A\&A
Rev. 8 279

\bibitem []  {}
Campana S., Stella L. 2000, ApJ 541 849

\bibitem []  {}
Campana S., Stella L., Mereghetti S., et al. 1998b, ApJ 499 L65 

\bibitem []  {}
Campana S., Gastaldello F., Mereghetti S., et al. 2002, ApJ 575 L15 

\bibitem []  {}
Campana S., Ravasio M., Israel G.L., Mangano V., Belloni T. 2003, ApJ 594 L39

\bibitem []  {}
Campana S., D'Avanzo, P., Casares J., et al. 2004, ApJ 601 474

\bibitem []  {}
Cash W. 1979, ApJ 228 939

\bibitem []  {}
Colpi M., Geppert U., Page D., Possenti A. 2001, ApJ 548 L175

\bibitem []  {}
Galloway D.K., Chakrabarty D., Morgan E.H., Remillard R.A., 2002, ApJ 576 L137

\bibitem []  {}
Gonz\'alez Hernandez J.I., Rebolo R., Pe\~narrubia J., Casares J.
Israelian G. 2005, A\&A
submitted (astro-ph/0502455) 

\bibitem []  {}
Heinke C.O., Grindlay J.E., Lugger P.M., et al. 2003, ApJ 598 501

\bibitem []  {}
Jonker P.G., Nelemans G. 2004, MNRAS 354 355

\bibitem []  {}
Jonker P.G., Galloway D.K., McClintock J.E., et al. 2004a, MNRAS 354 666 

\bibitem []  {}
Jonker P.G., Wijnands R., van der Klis M. 2004b, MNRAS 349 94

\bibitem []  {}
Jonker P.G., Campana S., Steeghs D., et al. 2005, ApJ submitted 

\bibitem []  {}
Lasota J.-P. 2001, NewAR 45 449

\bibitem []  {}
Markwardt C.B., Swank J.H., Strohmayer T.E., in't Zand J.J.M., Marshall
F.E. 2002, 575 L21

\bibitem []  {}
Markwardt C.B., Smith E., Swank J.H. 2003, IAUC 8080

\bibitem []  {}
Markwardt C.B., Juda M., Swank J.H. 2003, IAUC 8095

\bibitem []  {}
Markwardt C.B., Swank J.H., Strohmayer T.E. 2004, Atel 353

\bibitem []  {}
Remillard R.A., Swank J.H., Strohmayer T.E. 2002, IAUC 7893

\bibitem []  {}
Tomsick J.A., Gelino D.M., Halpern J.P., Kaaret P. 2004, ApJ 610 933

\bibitem []  {}
Wijnands R., van der Klis  M. 1998, Nat 394 344

\bibitem []  {}
Wijnands R., Miller J.M., Markwardt C., Lewin W.H.G., van der Klis M. 2001,
ApJ 560 L159 

\bibitem []  {}
Wijnands R., Homan J., Miller J.M., Lewin W.H.G. 2004, ApJ 606 L61

\bibitem []  {}
Wijnands R., Heinke C.O., Pooley D., et al. 2005a, ApJ 618 883

\bibitem []  {}
Wijnands R., Homan J., Heinke C.O., Miller J.M., Lewin W.H.G.	
2005b, ApJ 619 492

\bibitem []  {}
Yakovlev, D.G., Pethick, C.J. 2004, ARA\&A 42 169

\bibitem []  {}
in't Zand J.J.M., Cornelisse R., Kuulkers E., et al. 2001, A\&A 372 916


\end{thebibliography}
\end{document}